\documentclass[reprint,a4paper,twocolumn,nofootinbib,amsmath,amssymb,prd]{revtex4-1}
\pdfoutput=1

\usepackage[utf8x]{inputenc}
\usepackage{graphicx,hyperref}
\usepackage{float}

\newcommand{\neff}{n_{\rm eff}}
\renewcommand{\d}{{\rm d}}
\renewcommand{\bar}{\overline}
\newcommand{\hoppet}{\textsc{Hoppet}}
\newcommand{\R}[2]{$R_{#1#2}$}
\newcommand{\as}{$\alpha_s$}
\newcommand{\aw}{$\alpha_W$}
\newcommand{\nlojet}{\textsc{NLOJet++}}
\newcommand{\beq}{\begin{eqnarray}}
\newcommand{\eeq}{\end{eqnarray}}
\newcommand{\pTonetwo}{\left<p_{T1,2}\right>}

\begin{document}

\title{Constraining new colored matter from the ratio\\ of 3- to 2-jets cross sections at the LHC}

\author{Diego Becciolini}
\email{becciolini@cp3-origins.net}
\affiliation{$CP^3$--Origins \& DIAS, University of Southern Denmark, Campusvej 55, 5230 Odense M, Denmark}
\author{Marc Gillioz}
\email{gillioz@cp3-origins.net}
\affiliation{$CP^3$--Origins \& DIAS, University of Southern Denmark, Campusvej 55, 5230 Odense M, Denmark}
\author{Marco Nardecchia}
\email{m.nardecchia@damtp.cam.ac.uk}
\affiliation{DAMTP, University of Cambridge, Wilberforce Road, Cambridge CB3 0WA, United Kingdom}
\affiliation{Cavendish Laboratory, University of Cambridge, JJ Thomson Avenue, Cambridge CB3 0HE, United Kingdom}
\author{Francesco Sannino}
\email{sannino@cp3-origins.net}
\affiliation{$CP^3$--Origins \& DIAS, University of Southern Denmark, Campusvej 55, 5230 Odense M, Denmark}
\author{Michael Spannowsky}
\email{michael.spannowsky@durham.ac.uk}
\affiliation{Institute for Particle Physics Phenomenology, Department of Physics, Durham University, DH1 3LE, United Kingdom}

\preprint{
{\raggedleft
CP3-Origins-2014-010 DNRF90\\
DIAS-2014-10\\
IPPP/14/26\\
DCPT/14/52\\
}
}

\begin{abstract}
The Large Hadron Collider experiments are probing the evolution of the strong coupling \as\ up to the TeV scale. We show how the ratio of 3- to 2-jets cross sections is affected by the presence of new physics and argue that it can be used to place a model-independent bound on new particles carrying QCD color charge. The current data potentially constrains such states to be heavier than a few hundred GeVs.
\end{abstract}

\maketitle

\section{Introduction}

So far the most relevant result obtained from the Large Hadron Collider has been the discovery of the Higgs boson~\cite{Aad:2012tfa,Chatrchyan:2012ufa}, 
however non-Higgs analyses are also very valuable. For example in~\cite{Chatrchyan:2013txa} the first determination of the strong coupling $\alpha_s (M_Z)$ from measurements of momentum scales beyond 0.6 TeV was presented.
This determination has been performed studying the behaviour of the ratio \R32 of the inclusive 3-jet cross section 
to the 2-jet cross section, defined in greater detail in the next section. The result is in agreement with the world average value of $\alpha_s (M_Z)$.\footnote{Even more recent measurements of \as\ at high energy scales have appeared after completion of this project, see~\cite{Khachatryan:2014waa}.}

In this paper, we argue that it is in principle possible to constrain the presence of new colored states using such a measurement that probes quantum chromodynamics (QCD) at harder scales than ever before.
There are however some serious concerns regarding the validity of the interpretation given in the experimental analyses which warrants further studies.
We shall here focus on the potential value of such an observable for placing bounds on new physics beyond the Standard Model (BSM), granted that there is indeed a way to extract the value of the strong coupling constant at large momentum transfers from the data.
A review of the measurement procedure and of the observable itself is left for future work.
Instead, we offer some insights related to the presence of new (colored) particles. We show in particular that their effect on the parton distribution functions is negligible, at least when taking ratios of cross-sections. In the absence of striking final states, the presence of new colored particles must be tracked down to the running of \as.

The value of a good determination of \as\ lies in the fact that new colored states would modify its running regardless of their properties.
While a crafty model builder may be able to hide large numbers of particles under the current limits from direct searches by making them difficult to produce or see, if these carry color quantum numbers they may significantly contribute to QCD observables through virtual corrections.
This approach thus provides complementary information with respect to typical direct limits, where several assumptions have to be made in order to specify production and decay of a given particle.
For instance if the new particles have the right quantum numbers, searches for di-jet resonances are particularly constraining~\cite{Han:2010rf}, while there are models evading these bounds for which the results we present here may be relevant~\cite{Kubo:2014ova}.
Furthermore the effect on the running of \as\ only depends on the mass of the new states and on their color representation (and number), and an exclusion bound from such a measurement is, to a good approximation, model independent in that sense.

Efforts to constrain light colored states in the same spirit as the present work have already appeared. For example in~\cite{Berger:2004mj,Berger:2010rj} the authors considered the effect of a gluino-like state on the global analysis of scattering hadron data or in~\cite{Kaplan:2008pt} where model-independent bounds on new colored particles are derived using event shape data from the LEP experiments.
If the precision and energy reach claimed by the LHC collaborations do hold up to further scrutiny, we show that an improvement of nearly one order of magnitude can be derived with respect to these works.

Finally, this type of approach generalises to other sectors of the Standard Model, and the electroweak sector could for instance be constrained from high energy measurements of Drell-Yan~\cite{ruderman,Alves:2014cda}.

The paper is organized as follows: in Section 2 we present the observable \R32 and the main results obtained from~\cite{Chatrchyan:2013txa}, in Section 3 we analyse the effects of new physics, 
in Section 4 we present a series of expected bounds for different scenarios while in Section 5 we offer our conclusions.

\section{The \R32 observable}

Considering that we wish to test QCD at the highest possible energy scales, we are naturally interested in observables involving a low inclusive number of hard jets.
Furthermore, the best way to keep under control uncertainties --- both theoretical and experimental --- is to look at ratios.
The ideal candidate according to these criteria is the ratio of 3- to 2-jets (differential) cross sections, \R32, which is a commonly studied quantity~\cite{Arnison:1985zm, Appel:1985iv, Abe:1995rw, Abbott:2000ua, ATLAS:2013lla, Chatrchyan:2013txa}.

We focus on the following definition of the observable, in accordance with the latest CMS analysis~\cite{Chatrchyan:2013txa}:
\beq
	R_{32}\left( \pTonetwo \right) \equiv \frac{\d\sigma^{n_j\geq3} / \d\pTonetwo}{\d\sigma^{n_j\geq2} / \d\pTonetwo},
\eeq
where $\pTonetwo$ is the average transverse momentum of the two leading jets in the event,
\beq
	\pTonetwo \equiv \frac{p_{T1} + p_{T2}}{2}.
\eeq
Other choices are possible regarding the kinematic variable: one could use the $p_T$ of the leading jet only, or the sum of the $p_T$ of all the jets, or construct more complicated combinations as done for the observable $N_{32}$ considered by the ATLAS collaboration~\cite{ATLAS:2013lla}.

The CMS analysis we chose to follow is based on 5~fb$^{-1}$ of data collected at 7~TeV centre-of-mass energy~\cite{Chatrchyan:2013txa}.
Jets are defined requiring transverse momenta of at least 150~GeV and rapidities less than 2.5, using the anti-kT algorithm~\cite{Cacciari:2008gp} with size parameter R = 0.7 and E-recombination scheme.

The state-of-the-art computations for inclusive multijet cross sections include the next-to-leading order corrections in \as~and \aw~\cite{Ellis:1992en, Nagy:2001fj, Moretti:2006ea, Dittmaier:2012kx}.%
\footnote{Recent progress making use of new unitarity-based techniques will probably allow for complete NNLO results in a near future~\cite{Ridder:2013mf, Currie:2013dwa}.}
NLO QCD corrections are implemented in \nlojet~\cite{Nagy:2003tz}, that allows to evaluate the 3- and 2-jets cross sections at the parton-level within the Standard Model.

The factorisation and the renormalisation scales are identified with $\pTonetwo$ in the theoretical calculations presented by CMS.
This is where the problem lies: since 3-jet events
involve multiple scales, this simplified assignment may not represent the dynamics in play appropriately enough to allow a straightforward interpretation of the experimental data as a measurement of \as\ at $\pTonetwo$;
the observable may be mainly sensitive to the value of the strong coupling at some fixed lower scale.
Although the ideas we present here hinge on a resolution of this issue, finding the proper redefinition or reinterpretation of \R32 goes beyond the original scope of this paper, which only aims at encouraging a BSM reading of QCD results.

\begin{figure}
\centering
\includegraphics[width=0.95\linewidth]{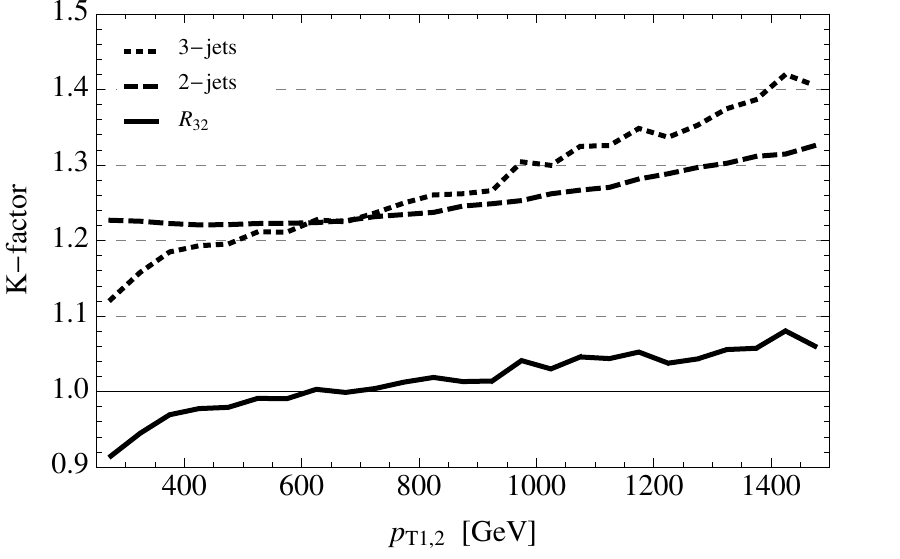}
\caption{The NLO $K$-factors of the 2- and 3-jets differential cross sections and of the ratio of the two, computed with \nlojet~\cite{Nagy:2003tz}, using the CTEQ~10 NNLO PDF set~\cite{Gao:2013xoa}.}
\label{fig:K_fact}
\end{figure}

As investigated in some detail in the experimental analyses,
the PDF uncertainty is reduced to few percent in the ratio;
in other words, the 3- and 2-jets cross section uncertainties are positively correlated, and there is no severe mismatch between the kinematic regions of the PDFs probed for a particular value of $\pTonetwo$ in the two cases~\cite{Nadolsky:2008zw}.
Other theoretical uncertainties, evaluated from the variation of the renormalisation, factorisation and resummation scales and tuning of the showering, are typically of order $5-10\%$ in the range of interest \cite{Hoche:2012wh,Badger:2013yda}.
Regardless of what the relevant scales in the process are and whether there are more appropriate kinematic choices for the observable itself, the sensitivity to higher order correction does seem to decrease in the ratio, as also seen from the reduction of the $K$-factor 
\beq
	K_{32}(\pTonetwo) \equiv \frac{R_{32}^{NLO}(\pTonetwo)}{R_{32}^{LO}(\pTonetwo)}
\eeq
compared to the ones of the individual differential cross sections, obtained from fixed-order computations, shown in Fig.~\ref{fig:K_fact}.

\begin{figure}
\centering
\includegraphics[width=0.95\linewidth]{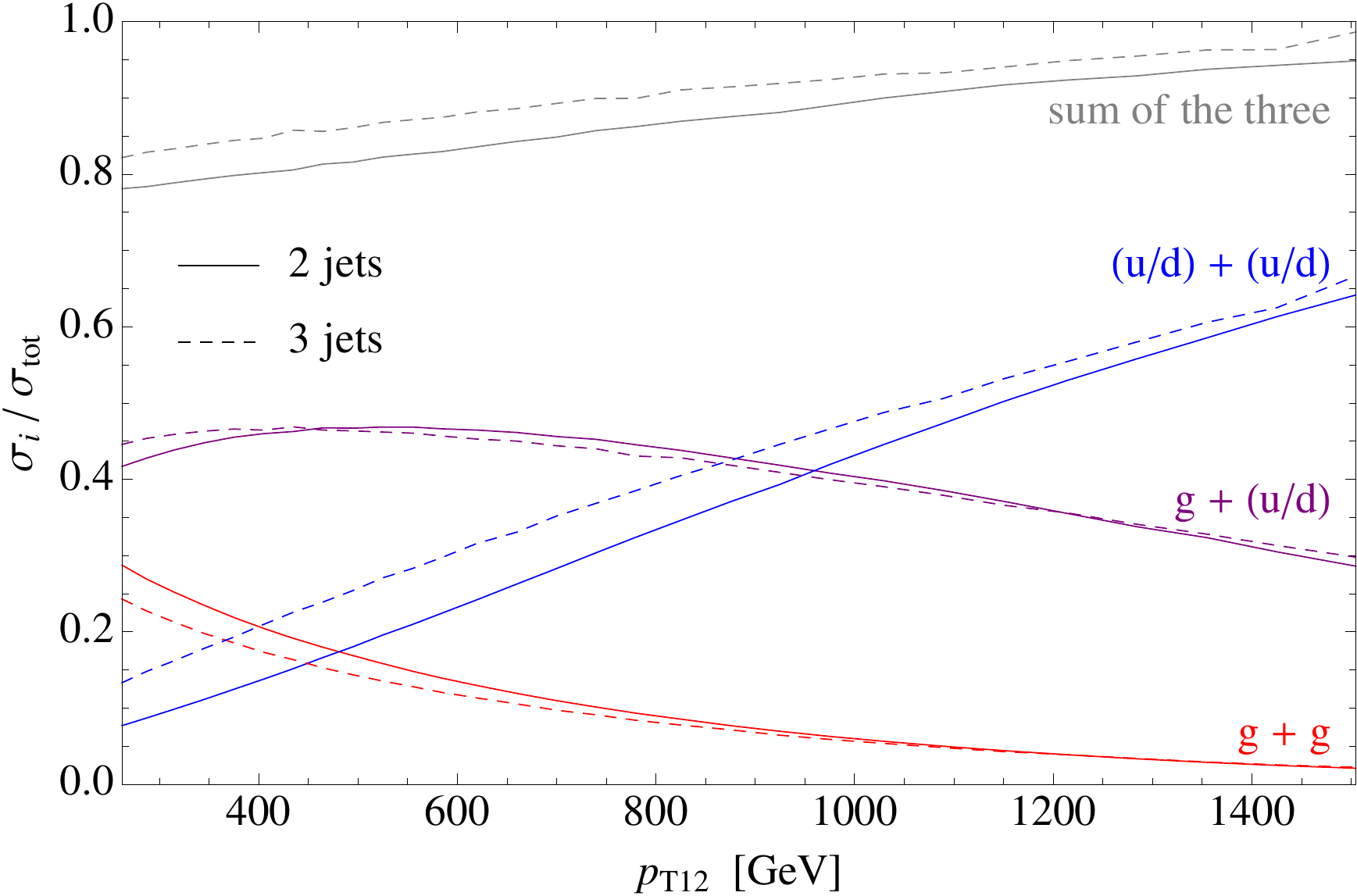}
\caption{Relative contributions of sub-processes at LO to 2- (continuous) and 3-jets (dashed) differential cross sections, selected according to the initial-state partons: only valence quarks (blue), one valence quark and one gluon (purple), two gluons (red), the sum of these three contributions (grey). All curves are obtained with \nlojet~\cite{Nagy:2003tz} and the CTEQ~10 NNLO PDF set~\cite{Gao:2013xoa}.}
\label{fig:subproc}
\end{figure}

We will see in the next section that the parton distribution functions of the gluon can be significantly affected by the presence of new colored particles. It is thus instructive to describe here the relative importance of the various sub-processes that enter the parton-level cross sections, using the Standard Model as a benchmark. At higher $p_T$, larger values of the momentum fraction of the proton are probed, where the dominant PDFs are the ones of the valence quarks (up- and down-quarks).
In Fig.~\ref{fig:subproc}, obtained again using \nlojet, we show that the processes with two valence quarks in the initial state become dominant indeed, while the second most important contribution comes from processes with one valence quark and one gluon.
The gluon fusion processes, as well as all other processes involving non-valence quarks, contribute little to hard events.

\begin{figure}
\centering
\includegraphics[width=0.95\linewidth]{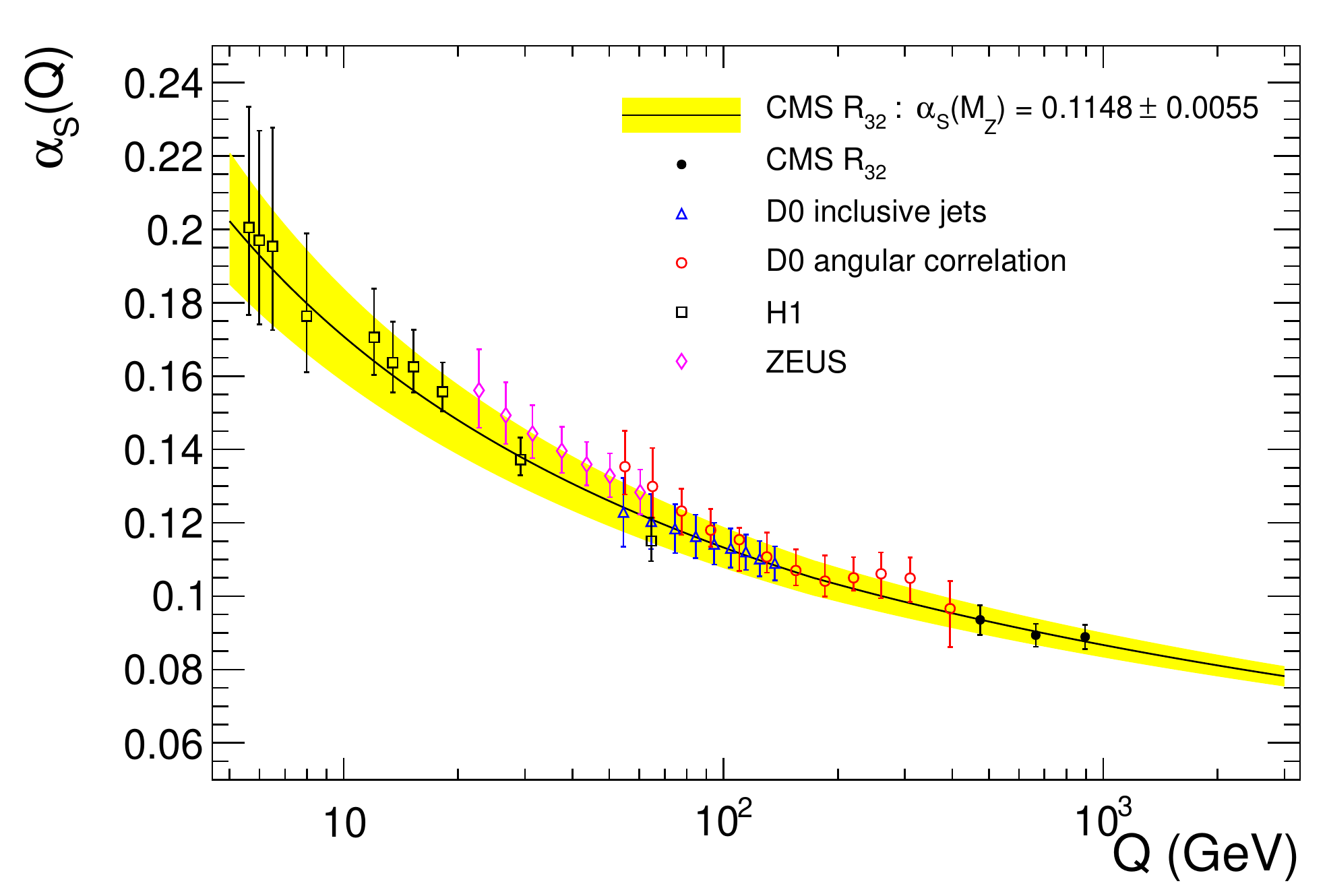}
\caption{\as\ measured at different energy scales and compared to the running obtained in the standard model. This figure is taken from~\cite{Chatrchyan:2013txa}.}
\label{fig:CMSas}
\end{figure}

Even though some experimental analyses do present the results of their fits in terms of the value of \as\ at the energy being probed (see Fig.~\ref{fig:CMSas}), the emphasis is on the results obtained by extrapolating back down to the usual reference scale $M_Z$, assuming the validity of the Standard Model (SM).
The final result in the latest CMS analysis is
\beq
    \alpha_s(M_Z) = 0.1148
    &\pm& 0.0014\,\text{(exp.)}
    \pm 0.0018\,\text{(PDF)} \nonumber\\
    &\pm& 0.0050\,\text{(theory)},
\eeq
and we see that the theoretical error is indeed the dominant one.
We take here a different approach and argue that bounds on potential BSM physics can be derived from such measurements.

\section{\R32 in the presence of new physics}

In this section, we discuss how hypothetical new colored particles can contribute to \R32.
This can happen through a modification of the running of \as\ and of the PDFs, and as additional contributions to the partonic cross section at leading or next-to-leading order.
We argue that the most important of these effects is the change in \as\ and that the correspondence between \R32 and  the strong coupling constant is reliable, even in the presence of new physics.

\subsection{New physics contribution to hard scattering}

In general, new states may contribute at tree-level to the jet cross sections if their quantum numbers allow it.
This would lead to important modifications, especially if there are resonant channels, and thus dedicated searches assuming specific production and decay mechanisms for the hypothetical particles are best suited to derive exclusion bounds.
In a ratio such as \R32, in fact, such contributions may in fact partially cancel, the same way NLO corrections do as shown in the previous section.
So when studying this observable, we are interested in cases where only virtual corrections would affect 2- and 3-jets cross sections, for instance in the case of fermions that do not mix with the SM quarks.

With our definition of \R32, jets in the final state are assumed to originate from a standard model parton, that is a quark or a gluon. The new colored fermions would of course also be produced copiously above the kinematic threshold, but we expect them to give a characteristic signature in the detectors. We do not study this case here, as it depends heavily on the couplings of the new fermions to standard matter, and is therefore model dependent. Notice that if it is stable, a new, heavy fermion in the final state could in principle be misidentified as a jet, since it would hadronize and end its path somewhere in the detector. There are however stringent constraints on the existence of such bound states~\cite{Aad:2013gva}.
The only remaining processes involving new physics at tree level are therefore those with a heavy fermion and a heavy antifermion in the initial state.
In spite of the possible enhancement of such processes due to large color factors, they remain negligible due to the minor importance of the PDF of the heavy fermion, as will be show in Section~\ref{sec:pdf}.

As discussed above, radiative corrections to \R32 are subdominant in the standard model, and so are they in the presence of new physics. The existence of loop diagrams involving new fermions introduces a new scale in the process at the loop level, and moreover the color factors can be enhanced in the case of fermions in a higher-dimensional representation; however, this dependence is made marginal in taking the ratio of 3- to 2-jets cross sections.
The only remaining contributions that would not obviously cancel in the ratio are threshold effects;
these have, to our knowledge, not been studied in 3-jets observables, but they do not modify 2-jets differential cross sections much~\cite{Ellis:1996sj,Ellis:1997iz}.
In the absence of an explicit NLO computation with massive fermions, our working assumption is that the correspondence between \R32\ and \as is not significantly affected by those effects.

\subsection{Running of \as}
\label{sec:runningalpha}

In the presence of new colored fermions, the running of \as\ at high energy is modified compared to the Standard Model, manifested by the introduction of new coefficients in the $\beta$ function. If we denote
\beq
	\beta(\alpha_s) \equiv \mu \frac{\partial \alpha_s}{\partial \mu} = - \frac{\alpha_s^2}{2 \pi} \left( b_0 + \frac{\alpha_s}{4 \pi} \, b_1 + \ldots \right),
\label{eq:beta}
\eeq
then the coefficients $b_0$ and $b_1$ in any mass-independent renormalisation scheme read
\beq
	b_0 & = & 11 - \frac{2}{3} n_f - \frac{4}{3} n_X T_X, \\
	b_1 & = & 102 - \frac{38}{3} n_f - 20 \, n_X T_X \left( 1 + \frac{C_X}{5} \right),
\eeq
where $n_f$ is the number of quark flavours (i.e. $n_f = 6$ at scales $Q > m_t$), $n_X$ the number of new (Dirac) fermions, and $T_X$ and $C_X$ group theoretical factors depending in which representation of the color group the new fermions transform. One has respectively for the fundamental (dimension 3), adjoint (8), two-index symmetric (6) and three-index symmetric (10) representations,
\beq
	{\renewcommand*{\arraystretch}{1.5}
	\begin{array}{l@{\hspace{0.5cm}}l@{\hspace{0.5cm}}l@{\hspace{0.5cm}}l}
		T_\mathbf{3} = \frac{1}{2}, &
		T_\mathbf{8} = 3, &
		T_\mathbf{6} = \frac{5}{2}, &
		T_\mathbf{10} = \frac{15}{2}, \\
		C_\mathbf{3} = \frac{4}{3}, &
		C_\mathbf{8} = 3, &
		C_\mathbf{6} = \frac{10}{3}, &
		C_\mathbf{10} = 6.
	\end{array}}
	\label{eq:reps}
\eeq
Higher-dimensional representations have typically larger values of $T_X$ and $C_X$, but will not be considered further in this work. The case of fermions in the adjoint representation --- like the gluino in the MSSM --- is special, since the representation is real: a Majorana mass term can be written for a single Weyl fermion, and $n_X$ can take half-integer values. At leading order, the modification in the running of \as\ only depends on a single parameter $\neff \equiv 2 n_X T_X$, counting the effective number of new fermions. Explicitly, we have
\beq
	\neff = n_{\mathbf{3}\oplus \bar{\mathbf{3}}} + 3 \, n_{\mathbf{8}}
		+ 5 \, n_{\mathbf{6}\oplus \bar{\mathbf{6}}}
		+ 15 \, n_{\mathbf{10}\oplus \bar{\mathbf{10}}},
	\label{eq:neff}
\eeq
where $n_{\mathbf{3}\oplus \bar{\mathbf{3}}}$, $n_{\mathbf{6}\oplus \bar{\mathbf{6}}}$ and $n_{\mathbf{10}\oplus \bar{\mathbf{10}}}$ are the number of new Dirac fermions in the triplet, sextet and decuplet representations respectively, and $n_{\mathbf{8}}$ the number of Weyl fermions in the adjoint representation. Asymptotic freedom is lost for $\neff > 10.5$. We do not restrict ourselves to asymptotically free theories.

Furthermore, one Dirac fermion corresponds to four complex scalar degrees of freedom;
scalar particles in the spectrum thus contribute to $\neff$ four times less than corresponding Dirac fermions.
For instance, the full content of the Minimal Supersymmetric Standard Model (1 adjoint Weyl fermion and 12 fundamental complex scalars) counts as $\neff = 6$.

Beyond leading order, $\neff$ is not sufficient to parametrise the effect of new physics and the detailed properties of the additional particles enter the computation.
Besides the value of the Casimir \eqref{eq:reps} also contributions from other sectors will influence the running of \as~\cite{Mihaila:2012fm}.
However these contributions are typically sub-leading and therefore a description in terms of $\neff$ is a useful approximation.

\begin{figure}
	\centering
	\includegraphics[width=0.95\linewidth]{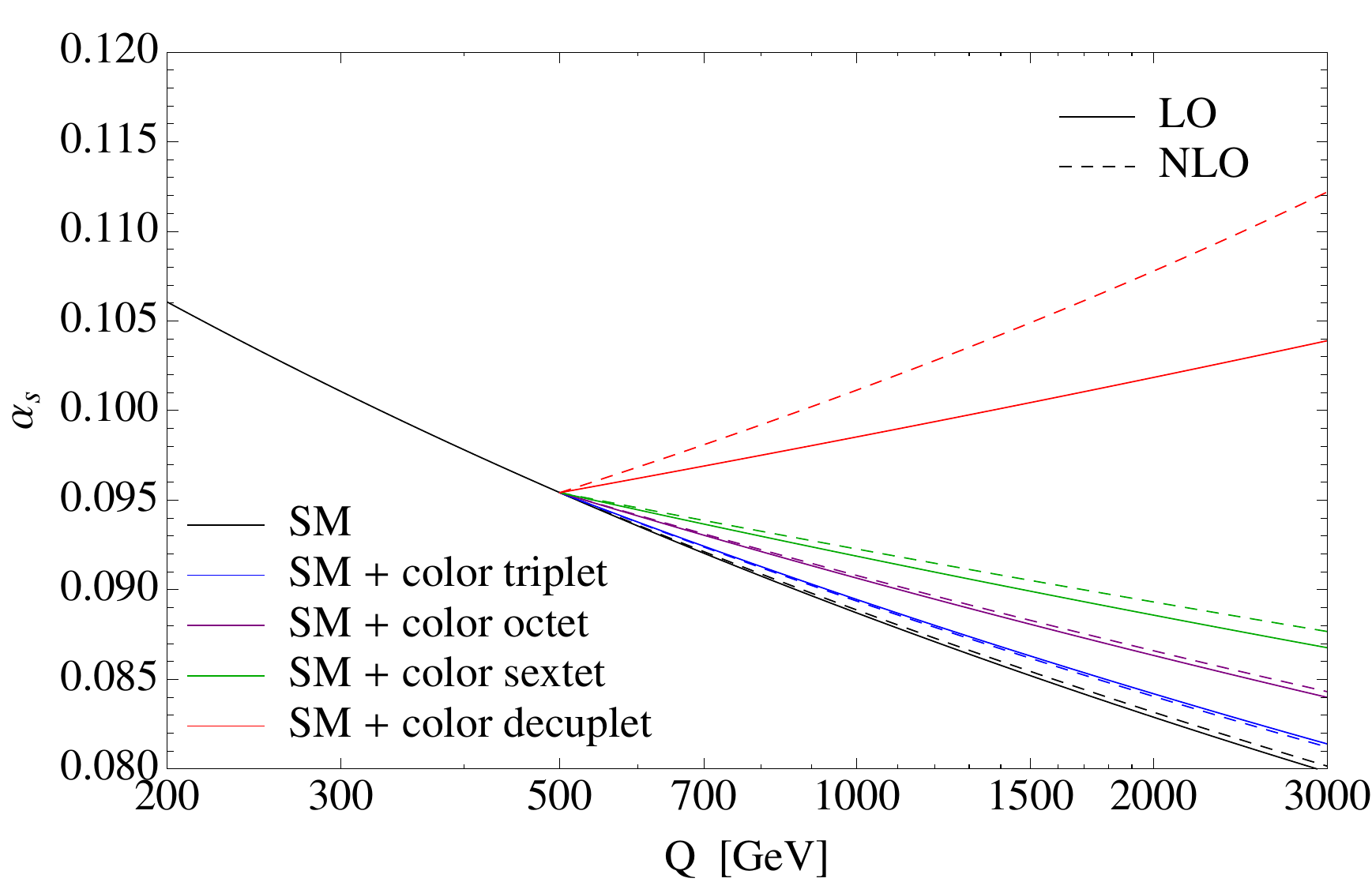}
	\caption{Example of the change in \as\ induced by a new fermion of mass 500~GeV in various representations of the color gauge group. The running of \as\ is performed at NLO, showing for comparison the running at LO from the mass of the new fermion.}
	\label{fig:alpha}
\end{figure}
The running of \as\ as given by the $\beta$ function above is only valid at energies larger than the mass of the new colored fermions --- for simplicity, we assume that they all have the same mass $m_X$ and that they are heavier than the top quark.
Following the standard procedure, we choose to perform the matching of \as\ between the high-energy regime and the effective theory without the new fermions exactly at the mass $m_X$. 
The choice of the matching scale $Q$ is arbitrary and the condition $Q=m_X$ is not in itself a requirement of the theory. However this choice leads to approximate continuity of the running coupling constant and hence 
a more appealing physical picture of \as\ (see for example~\cite{Ellis:1991qj}).

The relative importance of the change in \as\ induced by fermions in various representation can be assessed from Fig.~\ref{fig:alpha}.
LHC observables can only be sensitive to scales of a few TeV at most, and since we assume in our analysis that $m_X > m_t$, the modified running of \as\ will not take place over many orders of magnitude.
So with the exception of color-decuplet fermions, the modified running of \as\ is well estimated by the leading order running, which will allow us to provide model-independent bounds on new physics depending on $\neff$ and $m_X$ only.

We also derive the following approximate expression, which will be useful later in our discussion:
\beq
	\frac{\alpha_s(Q)}{\alpha_s^{SM}(Q)} \approx  1 + \frac{\neff}{3\pi} \alpha_s(m_X)
				\log\left( \frac{Q}{m_X} \right),
	\nonumber\\ \textrm{~for~} Q \geq m_X,
	\label{eq:alphaapprox}
\eeq
where $\alpha_s^{SM}(Q)$ is the Standard Model value of the running coupling.

\subsection{Parton distribution functions}
\label{sec:pdf}

New colored fermions also affect the parton distribution functions (PDFs), besides the QCD processes at the level of the parton interactions and through the modified running of \as. On one hand, their presence modifies the evolution of the PDFs of the quarks and gluons. On the other hand, they contribute as new partons to the momentum of the colliding protons. We will show here that in the case of \R32, the modifications of PDFs can actually be neglected. In order to assess the importance of new physics effects, we make use of a modified version of \hoppet~\cite{Salam:2008qg} to study the evolution of the PDFs above the scale of new physics. For simplicity, the evolution is performed using the DGLAP equations at leading order only and then compared to the PDFs in the Standard Model evaluated at the same order. This gives a good estimate of the modifications induced by the new fermions.
The explicit procedure followed is to initialise the PDFs at the scale $Q = m_t$ with the CTEQ distribution~\cite{Gao:2013xoa}, then use the DGLAP equations~\eqref{eq:DGLAP} of Appendix~\ref{sec:DGLAP} to perform the evolution above this energy.%
\footnote{In principle we could use the SM PDFs up to the mass $m_X$ of the new fermions. However, the CTEQ PDF set (like most others sets) does not make use of the 6 flavours running scheme above the top mass.}
The results can be summarised in three points:
\begin{enumerate}

\item
The evolution of the PDF of the new fermions above the mass threshold is driven by the gluon PDF, and it is therefore proportional to the splitting function $P_{Xg} \propto T_X$. Fermions in low-dimensional representations will therefore have a small PDF, at most comparable to that of the top quark. Fermions in a higher-dimensional representation of the color group will have PDFs increasing faster with energy, yet they remain small compared to the valence quarks and gluon PDFs over a large range of energies. As an example, Fig.~\ref{fig:pdfdirect} shows the PDFs of a new fermion of mass $m_X = 500$~GeV in the octet (left panel) and decuplet (right panel) representations, at factorisation scale $Q = 1.5$~TeV. Although the PDF of the new fermion becomes as important as that of the light quarks, it is still one or two order of magnitude below the relevant PDFs, i.e.~the valence quarks and/or the gluon ones depending on which kinematic region is considered.
\begin{figure*}
	\centering
	\includegraphics[width=0.49\linewidth]{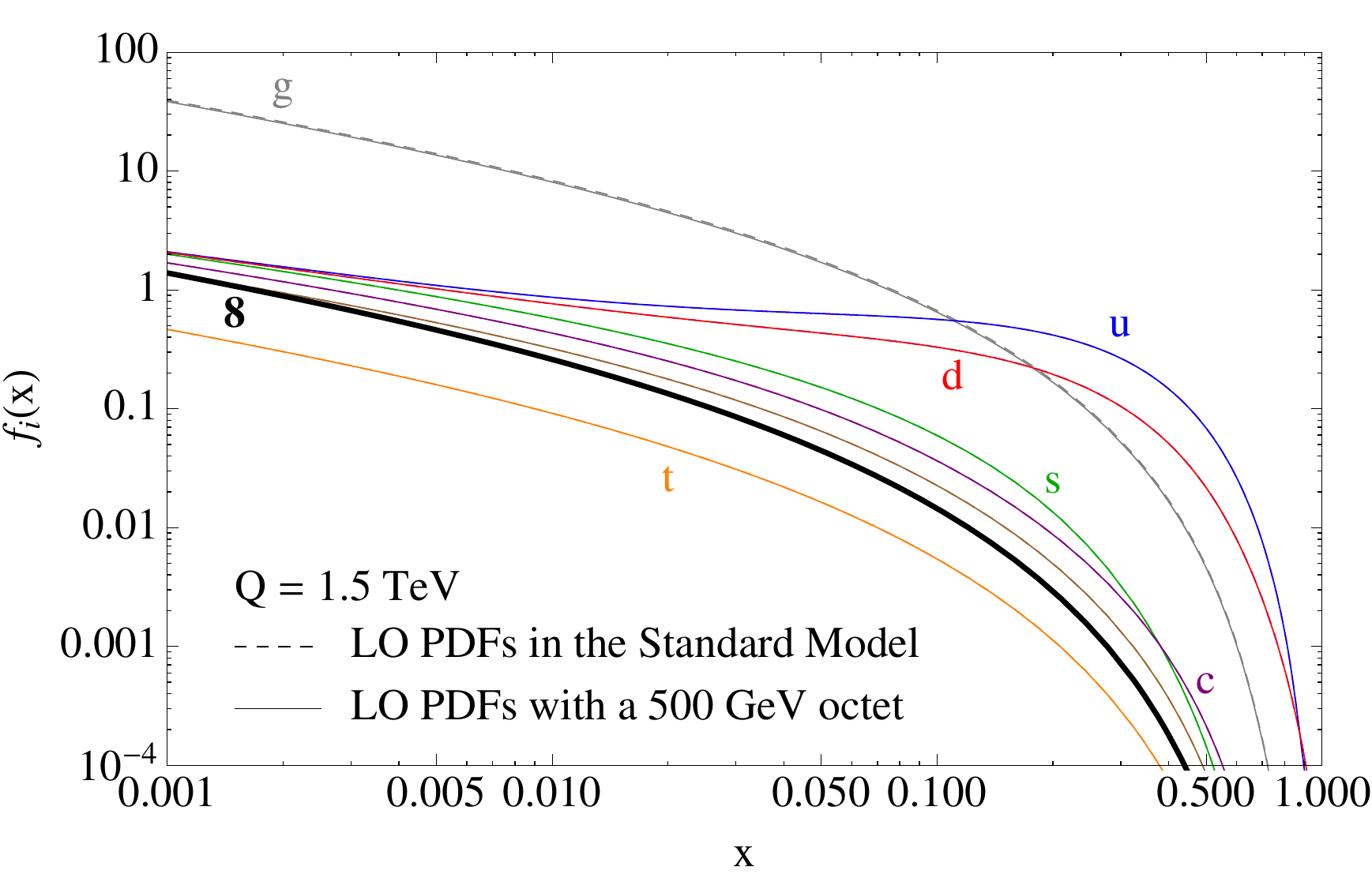}
	\includegraphics[width=0.49\linewidth]{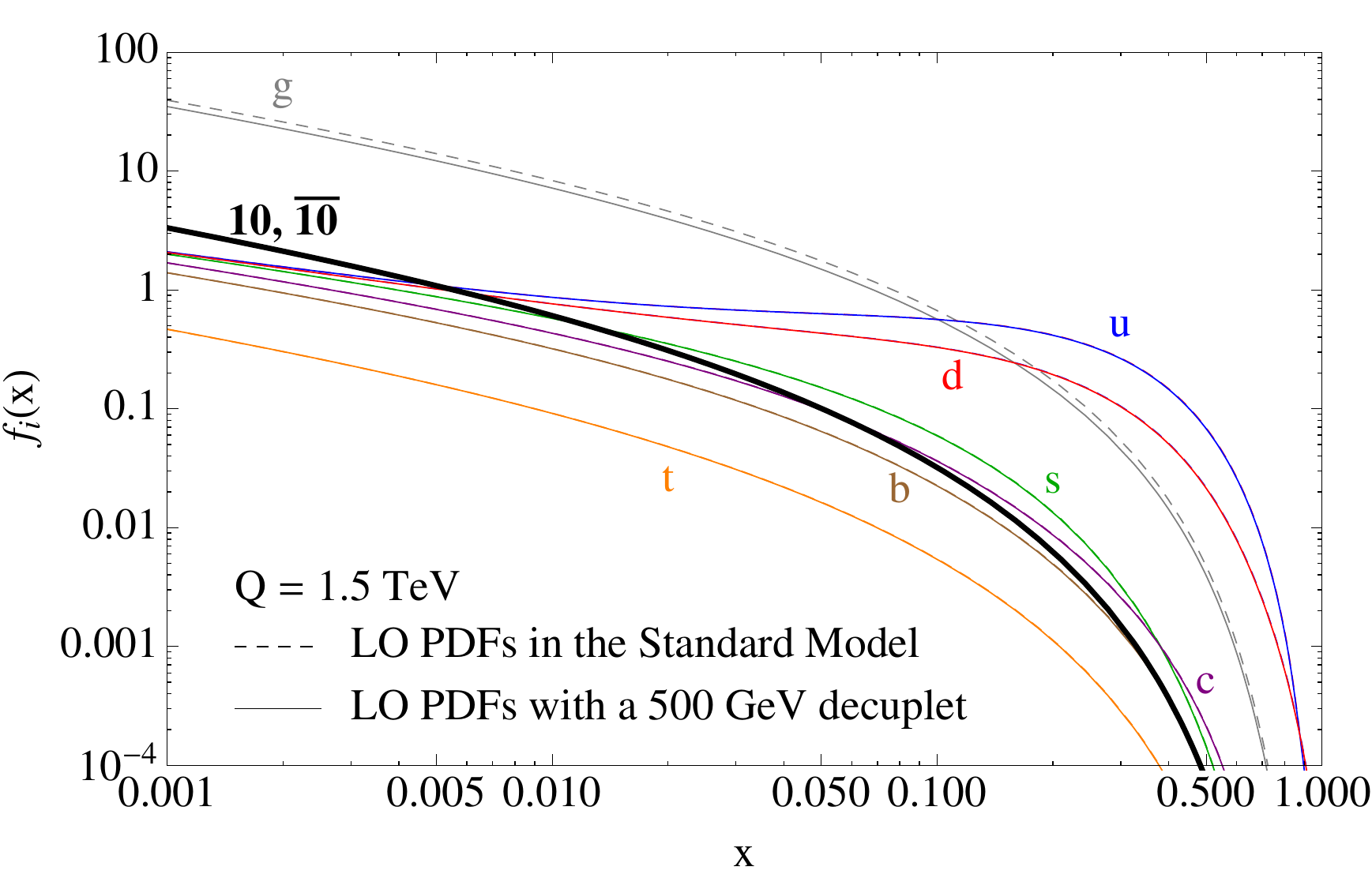}
	\caption{Relative importance of the PDFs with a new colored fermion of mass 500~GeV in the adjoint (left panel) or three-index symmetric representation (right panel). The PDF are shown at the scale $Q = 1.5$~TeV. The Standard Model PDFs are shown in comparison as dashed lines (not visible on the left panel since the discrepancy is tiny).}
	\label{fig:pdfdirect}
\end{figure*}

\item
The evolution of the gluon PDF is also largely dominated by the gluon PDF itself.
New physics enters therefore the DGLAP equations in two ways: in the value of \as\ above the mass threshold and in the splitting function of the gluon into itself, eq.~\eqref{eq:gluonsplittingfunction}. The former effect is actually subleading in \as, since the ratio $\alpha_s(Q)/\alpha_s^{SM}(Q)-1$ is itself proportional to \as, see eq.~\eqref{eq:alphaapprox}. The modified splitting function of the gluon, on the contrary, gives an important contribution to the evolution of the gluon PDF. It can be quantified by looking at the evolution of the ratio of the gluon PDF in the presence of new fermions to its counterpart in the Standard Model. One finds, at leading order in \as,
\beq
	Q^2 \frac{\partial}{\partial Q^2} \left(\frac{f_g(x,Q)}{f_g^{SM}(x,Q)}\right) && \nonumber\\
		\approx - \frac{\neff}{6\pi} &\alpha_s(Q)& \frac{f_g(x,Q)}{f_g^{SM}(x,Q)},
\eeq
which can be integrated to give
\beq
	\frac{f_g(x,Q)}{f_g^{SM}(x,Q)} \approx 1 - \frac{\neff}{3\pi} \alpha_s(m_X)
				\log\left( \frac{Q}{m_X} \right).
	\label{eq:gluonPDFreduction}
\eeq
The main effect of the presence of new fermions is therefore the reduction of the gluon PDF by a factor proportional to $\neff$ but independent of $x$. This behaviour is confirmed by the explicit evolution obtained with \hoppet. The left-hand side of Fig.~\ref{fig:pdfindirect} shows the leading order evolution obtained for three different cases all corresponding to $\neff = 15$, where all gluon PDFs are normalised to the Standard Model. Eq.~\eqref{eq:gluonPDFreduction} gives in this case $f_g(x,Q) \approx 0.83 \, f_g^{SM}(x,Q)$, which is indeed the behaviour observed.
\begin{figure*}
	\centering
	\includegraphics[width=0.49\linewidth]{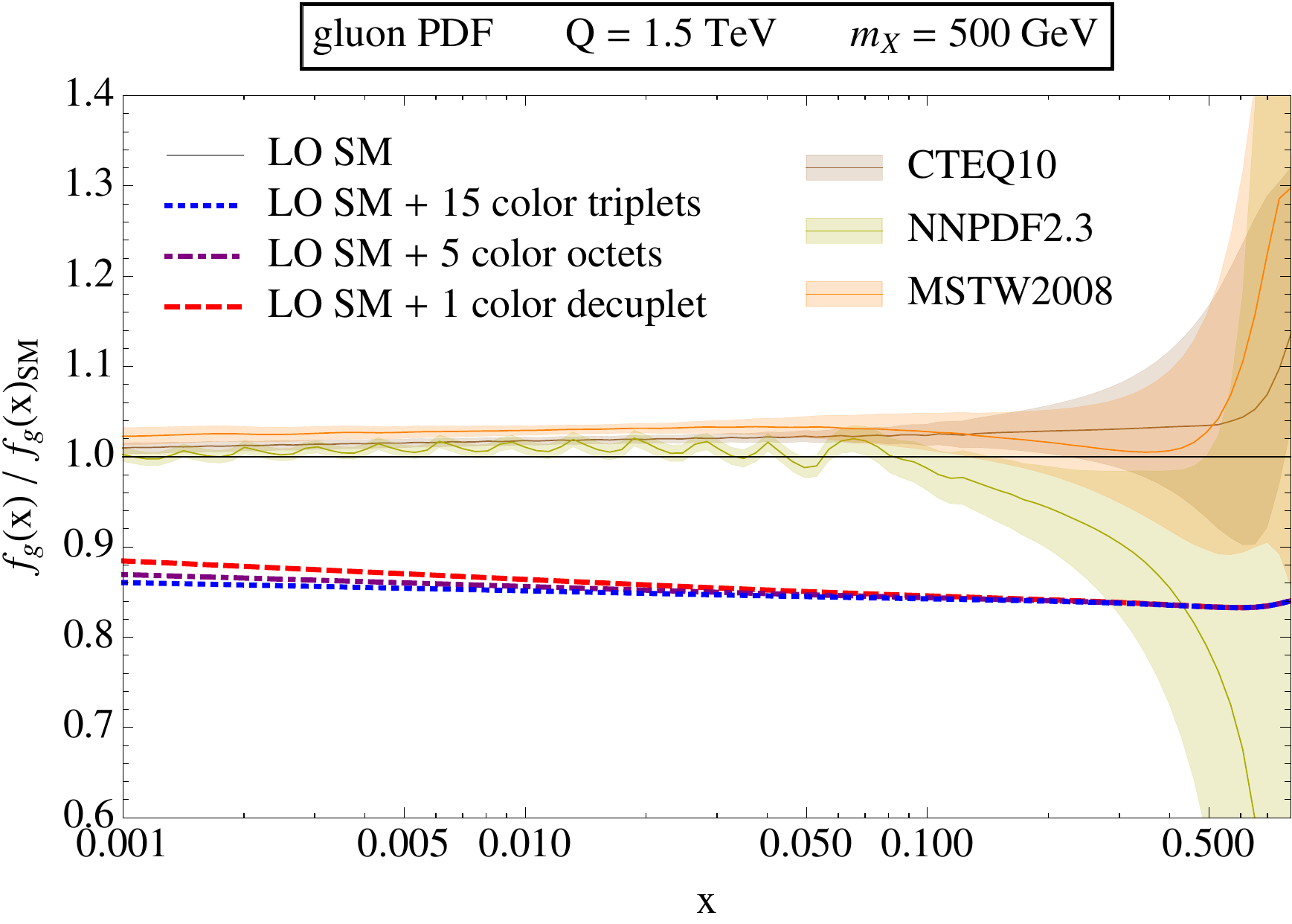}
	\includegraphics[width=0.49\linewidth]{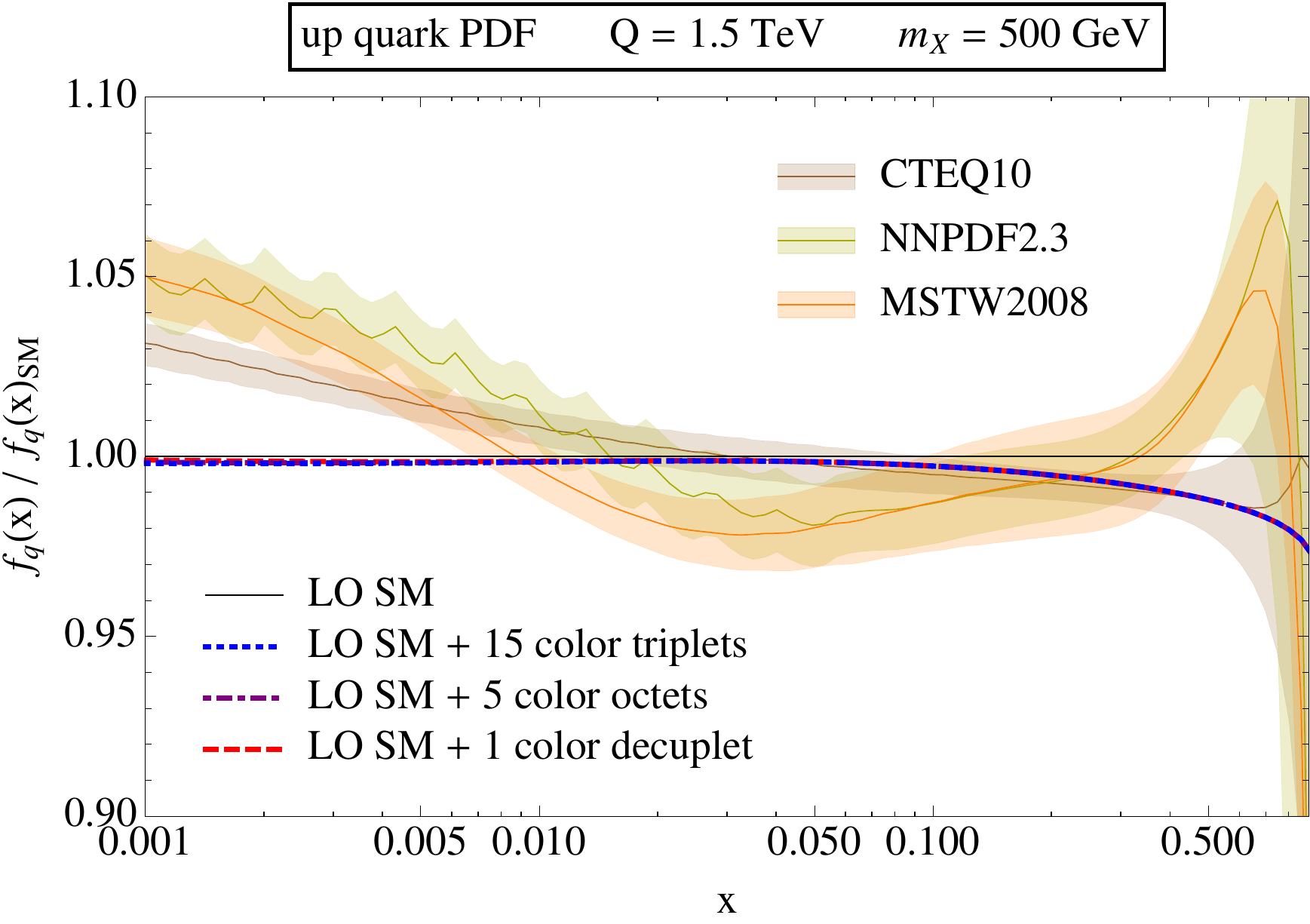}
	\caption{Change in the PDF of the gluon (left panel) and of the  up quark (right panel) at $Q = 1.5$~TeV induced by a large effective number of new fermions $\neff = 15$.  All PDFs are normalised to the Standard Model PDFs computed at LO, which explains the discrepancy with some of the usual PDF sets shown for comparison: CTEQ10 NNLO~\cite{Gao:2013xoa}, NNPDF 2.3~\cite{Ball:2012cx} and MSTW2008~\cite{Martin:2009iq}.}
	\label{fig:pdfindirect}
\end{figure*}

\item
The evolution of the quark PDF is less affected than the gluon PDF, for the simple reason that the splitting functions of the quark PDFs do not depend on the presence of new fermions. Moreover, it is indirectly feeling their presence through the modification of \as\ and of the gluon PDF, but the two corrections eqs.~\eqref{eq:alphaapprox} and~\eqref{eq:gluonPDFreduction} are equal and in opposite direction, so that they conspire and make the quark PDF mostly insensitive to new physics. Notice however that at large $x$ the PDF of the valence quarks are important as well in the evolution, and for them the enhancement in \as\ is not compensated, hence effectively accelerating the evolution, in this case reducing the quark PDFs at large $x$. This is very well visible   in the right-hand side of Fig.~\ref{fig:pdfindirect}, where we show the normalised PDFs of the up quark for three scenarios corresponding to $\neff = 15$. It should be noted in this case that the leading order evolution of the PDFs does not match very well the standard sets using the NNLO evolution equations; however, our point here is simply to show that the quark PDFs are affected very little by the presence of new fermions, which is clear from the figure.

\end{enumerate}
In general, the sizeable change in the gluon PDF could have important effects on physical observables at high factorisation scale $Q$. The ratio \R32, however, is barely sensitive to this change, since the relative contribution of the gluon-induced processes to the two and three jets differential cross section is identical, as seen in Fig.~\ref{fig:subproc}, and since the reduction of the gluon PDF is $x$-independent. At the precision level of our analysis, the modification of PDFs can thus be safely neglected.

\section{Bounds on the new physics}

In the previous sections, we argued that we expect hypothetical new colored physics to affect \R32 principally through a modification of the running of \as, implying that this observable could provide a robust determination of the strong coupling constant, at least if problems regarding the separation of scales in 3-jet events is resolved.
To illustrate the exclusion potential of high-scale measurements of \as\ we present bounds on $\neff$ depending on the scale of new physics $m_X$, derived from results provided by CMS~\cite{Chatrchyan:2013txa}.

\begin{table}
\centering
\begin{tabular}{c|c}
$Q$ [GeV] & $\alpha_s^{exp}(Q) \pm \sigma(Q)$ \\
\hline
474 & $0.0936 \pm 0.0041$ \\
664 & $0.0894 \pm 0.0031$ \\
896 & $0.0889 \pm 0.0034$ \\
\end{tabular}
\caption{High-scale determinations of \as\ from measurements of \R32 by CMS~\cite{Chatrchyan:2013txa}.}
\label{tab:alpha}
\end{table}
As our goal is mainly to encourage experimental groups to also interpret their measurements in terms of exclusion bounds on new physics, we chose to perform a simplistic analysis here as a proof of concept and simply take the estimates of \as\ given by CMS (and reproduced in Tab.~\ref{tab:alpha}) at face value.
We add to the analysis the the world average measurement of the strong coupling $\alpha_s(M_Z) = 0.1185 \pm 0.0006$~\cite{Beringer:1900zz}; since its uncertainty is much smaller than the ones of the other data points,
we take as fixed input $\alpha_s(M_Z) = 0.1185$.

We also simply assume the uncertainties to be Gaussian and these measurements to be independent.
The induced probability measure over the parameter-space we want to constrain is then simply proportional to
\beq
\exp\left[{-\frac{1}{2} \sum_Q \left(\frac{\alpha_s^{exp}(Q) - \alpha_s^{th}(Q ; \neff, m_X)}{\sigma(Q)}\right)^2}\right] \nonumber\\
\times \text{priors}, \quad
\eeq
where $\alpha_s^{th}(Q ; \neff, m_X)$ is the theoretical prediction for the value of the strong coupling at the scale $Q$, which is a function of $\neff$ and $m_X$.

The theoretical predictions for \as\ are obtained by running up to $Q$ from the $Z$-mass at two-loop order, as described in eq.~\eqref{eq:beta}, which is sufficient for our purpose.
Beyond leading-order, $\neff$ is not enough to parametrise the importance of new physics effects:
the quadratic Casimir $C_X$ needs to be specified.
We vary it between $4/3$ and $6$ --- the values corresponding to fermions in the fundamental or decuplet representations, respectively --- to show that, as a subleading effect, it has little influence.

\begin{figure}
	\centering
		\includegraphics[width=0.95\linewidth]{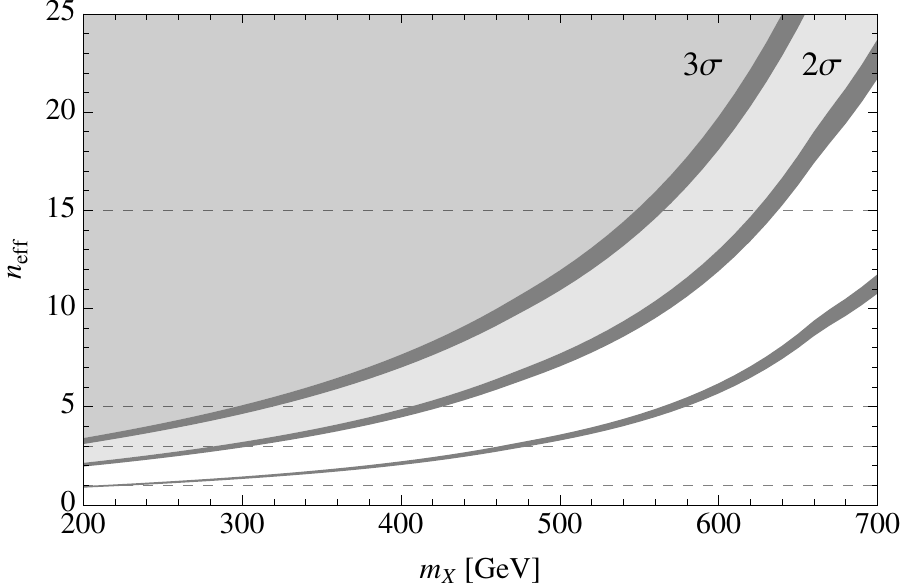}
	\caption{The shaded regions indicate the upper bounds on $\neff$ at $2\sigma$ and $3\sigma$ confidence levels, assuming the scale of new physics $m_X$ is known.
	They are delimited by grey bands whose width show the effect of varying the Casimir $C_X$.
	 As further indication, the third band shows a $1\sigma$ limit.
	To guide the eye, the dashed horizontal lines indicate values of $\neff$ corresponding to one fundamental, one adjoint, one two-index symmetric and one three-index symmetric fermion (see eq.~\eqref{eq:neff}).}
\label{fig:exclusion}
\end{figure}

A detailed interpretation of the uncertainty on the scale of new physics $m_X$ is beyond the scope of this analysis.
We thus assume that the mass of the new states is known and show the subsequent upper bound on $\neff$, choosing a flat prior over $\neff > 0$, in Fig.~\ref{fig:exclusion}.

Next year, the LHC will start its second run at circa double the centre-of-mass energy, finally attaining its original target.
This also means a doubling of the reach in the search for new physics:
the main factors determining the number of events occurring at a given scale are the steeply falling PDFs, thus the corresponding value of the momentum fraction.
Up to changes due to logarithmic scale corrections and modified experimental conditions, a same amount of data at twice the centre-of-mass energy would translate in mass-exclusion bounds roughly twice as high.
Of course, all searches will see their potential increase.
The relative simplicity of the analysis we suggest here may allow to extract limits on new physics rapidly as new data becomes available, again, provided that the theoretical footing of the observable can be established more firmly.

\section{Conclusions}

One should not overlook pure QCD observables as a means of placing bounds on new physics beyond the Standard Model.
Such bounds can indeed be insensitive to the detailed properties of the hypothetical states, as their various charges, and mainly depend on their effective number $\neff$ (and their mass).
These limits on colored particles, although not the most stringent for any specific model in general, would be largely unavoidable due to their model-independent nature.

We argue that the ratio of 3- to 2-jets inclusive differential cross sections \R32 would be particularly appropriate for constraining new physics since PDF uncertainties are suppressed and the main effect of additional heavy particles is encoded in the modified running of the strong coupling \as,
but we are unable at this point to resolve some issues regarding the proper interpretation of the observable.

We want to encourage on the one hand experimental collaborations to interpret their results not only as a test of the Standard Model, but also more directly as a probe of new physics, and on the other hand theorists to put their efforts in computing the relevant processes at higher orders in perturbation theory. Robust bounds on New Physics can only be derived with more work, first and foremost understanding precisely which scale is being probed in \R32.

If \as\ can indeed be measured to the precision currently estimated by experimental collaborations, based on the simplified analysis of Section 4 the exclusion potential of currently available experimental data is shown in Tab.~\ref{tab:limits}.

\begin{table}
\centering
\begin{tabular}{c|c|c}
 color content & $\neff$ & $m_X$ in GeV \\
\hline
Gluino & 3 & 280 \\
Dirac sextet & 5 & 410 \\
MSSM & 6  & 450 \\
Dirac decuplet & 15 & 620 \\
\end{tabular}
\caption{
$95\%$ CL mass exclusion bounds for various values of $\neff$ according to a toy-analysis of the latest CMS measurement of \R32~\cite{Chatrchyan:2013txa}.
}
\label{tab:limits}
\end{table}

\acknowledgments
We would like to thank Roberto Franceschini, Tuomas Hapola, Petar Petrov, Michele Redi, Marek Schoenherr and the members of the Cambridge SUSY Working Group for discussions and helpful comments, as well as Olivier Mattelaer for his help with MadGraph.
The CP$^3$-Origins centre is partially funded by the Danish National Research Foundation, grant number DNRF90.

\appendix
\begin{widetext}
\section{DGLAP equations with new colored fermions}
\label{sec:DGLAP}

Defining the sum of the PDF of all quarks as
\beq
	\Sigma(x, Q^2) = \sum_{q=q_i, \bar{q}_i} f_q(x,Q^2)
\eeq
and denoting by $f_X(x, Q^2)$ the PDF of the new colored fermion --- or the sum of them if they are more than one --- the DGLAP evolution equations are at leading order
\beq
	Q^2 \frac{\partial f_g}{\partial Q^2} (x,Q^2)
		& = & \frac{\alpha_s}{2 \pi} \int_x^1 \frac{\d z}{z}
		\left[ P_{gg}(z) f_g\left(\frac{x}{z}, Q^2\right)
		+ P_{gq}(z) \Sigma\left(\frac{x}{z}, Q^2\right)
		+ n_X(Q) P_{gX}(z) f_X\left(\frac{x}{z}, Q^2\right) \right],
		\nonumber\\
	Q^2 \frac{\partial \Sigma}{\partial Q^2} (x,Q^2)
		& = & \frac{\alpha_s}{2 \pi} \int_x^1 \frac{\d z}{z}
		\left[ P_{qq}(z) \Sigma\left(\frac{x}{z}, Q^2\right)
		+ 2 n_f(Q) P_{qg}(z) f_g\left(\frac{x}{z}, Q^2\right) \right],
	\nonumber \\
	Q^2 \frac{\partial f_X}{\partial Q^2} (x,Q^2)
		& = & \frac{\alpha_s}{2 \pi} \int_x^1 \frac{\d z}{z}
		\left[ P_{XX}(z) f_X\left(\frac{x}{z}, Q^2\right)
		+\frac{n_X(Q)}{n_X} P_{Xg}(z) f_g\left(\frac{x}{z}, Q^2\right) \right],
\label{eq:DGLAP} 		
\eeq
where $n_f(Q)$ and $n_X(Q)$ are the number of active flavours of quarks and new fermions respectively at scale $Q$. The splitting functions $P_{ij}$ are defined as
\beq
	P_{gg}(z) = 6 \left[ \frac{z}{(1-z)_+} + \frac{1-z}{z} + z (1-z) \right]
		&+& \frac{1}{2} \left(11 - \frac{2}{3} n_f(Q) - \frac{4}{3} n_X(Q) T_X \right)
		\delta(1-z),
	\label{eq:gluonsplittingfunction}
\\
	P_{qq}(z) = 3 \left( \frac{1+z^2}{1-z} \right)_+,&& \hspace{2cm}
	P_{XX}(z) = C_X \left( \frac{1+z^2}{1-z} \right)_+,
\\
	P_{gq}(z) = 3 \frac{1 + (1-z)^2}{z},&& \hspace{2cm}
	P_{gX}(z) = C_X \frac{1 + (1-z)^2}{z},
\\
	P_{qg}(z) = \frac{1}{2} \left[ z^2 + (1-z)^2 \right],&& \hspace{2cm}
	P_{Xg}(z) = T_X \left[ z^2 + (1-z)^2 \right],
	\label{eq:splittingfunctions}
\eeq
where the group invariants $T_X$ and $C_X$ are defined in Section~\ref{sec:runningalpha}.
\pagebreak
\end{widetext}

\bibliography{biblio}

\end{document}